# Composition-dependent bulk properties of intercalated transition metal dichalcogenides $Co_{1/3(1\pm\delta)}NbS_2$

Woonghee Cho,[1, 2*] Kiwan Nam,[2, 3*] Yeochan An,[1, 2] You Young Kim,[4] Myung-Hwa Jung,[4] Kee Hoon Kim,[2, 3] and Je-Geun Park[1, 2, 5, #]

[1]Center for Quantum Materials, Seoul National University, Seoul 08826, Republic of Korea
[2]Department of Physics and Astronomy, Seoul National University, Seoul 08826, Republic of Korea
[3]Center for Novel States of Complex Materials Research, Seoul National University, Seoul 08826, Republic of Korea
[4]Department of Physics, Sogang University, Seoul, Korea
[5]Institute of Applied Physics, Seoul National University, Seoul 08826, Republic of Korea

* Equal contribution
# Corresponding author: jgpark10@snu.ac.kr

## Abstract

We report a systematic study of the composition-dependent bulk properties in $Co_{1/3(1\pm\delta)}NbS_2$ single crystals across a series of precisely controlled cobalt compositions with -4% ≤ δ ≤ 8%. By tuning the cobalt stoichiometry, we find that the topological Hall effect is critically sensitive to the intercalant cobalt composition and is completely suppressed when the cobalt composition exceeds δ=+4%. We observe that the longitudinal conductivity is also strongly influenced by the cobalt composition, reaching its maximum value just before the disappearance of the topological Hall effect. Furthermore, heat capacity measurements reveal distinct Sommerfeld coefficients ($\gamma$) across different compositions, which exhibit a clear linear scaling with the inverse of the ordinary Hall coefficient ($R_H^{-1}$). These results demonstrate that composition tuning in $Co_{1/3(1\pm\delta)}NbS_2$ systematically modifies the low-energy electronic degree of freedom, moving beyond a simple dilute impurity picture. Finally, we use the microscopic spin Hamiltonian to explain the stability of the experimentally observed M-point modulation vector and the corresponding triple-**Q** magnetic order. Our findings highlight that the topological properties of this system are highly tunable through precise control of the intercalant concentration, offering a new perspective on the competition between electronic and magnetic orders in intercalated transition-metal dichalcogenides.

## Introduction

Two-dimensional (2D) magnetism has long been an important topic in modern condensed matter physics, beginning with the early theoretical studies of the Ising [1], XY [2,3], and Heisenberg [4] models. The recent discovery of van der Waals (vdW) magnets [5–7] has enabled many exciting, otherwise inaccessible, experimental studies using real magnetic systems that can be thinned down to the atomic limit. However, the number of well-characterized 2D magnetic materials remains relatively limited. $TMPS_3$ (TM = $3d$-transition metal) is one of the very good systems, with a diverse choice of TM elements, serving as a new representative example [5,8]. In this context, $3d$-transition-metal intercalated transition-metal dichalcogenides (TMDs), such as $TM_{1/3}TaS_2$ and $TM_{1/3}NbS_2$ (TM = $3d$ transition metal), first reported in the early 1980s [9], have recently attracted renewed interest for studies of 2D magnetism. In these systems, localized moments from the intercalated $3d$ transition metals couple to a metallic host through itinerant Ta/Nb $d$-electrons. This allows one to examine magnetic interactions mediated by itinerant electrons, in contrast to the short-range super-exchange interactions typical of insulating vdW magnets [10]. Several studies have investigated the coupling between itinerant electrons and localized moments in these systems [11–14].

Among the possible intercalant concentrations, the x = 1/3 composition is particularly interesting, as several of these compositions display diverse ground states. In this case, magnetic ions form a commensurate $\sqrt{3} \times \sqrt{3}$ triangular lattice with respect to the original TMD structure. At this concentration, the spacing between neighboring $3d$ transition metals exceeds the Hill limit [15], effectively suppressing direct exchange interactions. As a result, the dominant magnetic interaction becomes the Ruderman–Kittel–Kasuya–Yosida (RKKY) interaction [17–19], an indirect exchange mechanism mediated by itinerant electrons. The triangular lattice geometry further introduces geometric frustration, allowing several magnetic structures to be stabilized [16–20]. For instance, recent studies have reported non-coplanar triple-**Q** magnetic structures in $Co_{1/3}TaS_2$ and $Co_{1/3}NbS_2$, together with the associated topological Hall effect (THE) [18,20–23].

Generally, distinguishing a multi-**Q** phase from a multi-domain single-**Q** configuration is difficult in conventional diffraction due to spatial averaging. To resolve this ambiguity, investigating transport properties, particularly symmetry-sensitive responses, has served as a powerful complementary probe for multi-**Q** spin textures. For example, the M-point modulation vector has been identified with a spontaneous Hall effect alongside their magnetic ordering in $Co_{1/3}TaS_2$ and $Co_{1/3}NbS_2$ [21,23]. When considering the possible single-**Q** magnetic structures compatible with this known modulation vector, symmetry analysis shows that a collinear single-**Q** configuration at the M-point strictly preserves the combined time-reversal and fractional lattice translation symmetry ($\tau_a \circ T$). This constraint forbids a macroscopic Hall response ($\sigma_{xy} = 0$) for individual domains. To reconcile the M-point vector with the simultaneous emergence of the spontaneous Hall effect, it becomes compelling to invoke a non-coplanar triple-**Q** configuration, the atomistic limit of a skyrmion lattice, where the coherent superposition of multiple modulation vectors generates a non-vanishing scalar spin chirality.

Crucially, however, the stabilization of such non-coplanar spin textures and the resulting transport signatures heavily rely on the detailed intercalant stoichiometry. Indeed, magnetic and transport properties of several intercalated TMDs are known to be sensitive to small deviations from nominal intercalant concentration. In $Fe_{1/3(1\pm\delta)}NbS_2$ and $Co_{1/3(1\pm\delta)}TaS_2$, even small deviations from stoichiometry can significantly modify the electronic and magnetic ground states and their associated functionalities [24–27]. Interestingly, $Co_xNbS_2$ near x = 1/3 also shows strong sample dependence in its THE and bulk properties, as reported in different studies [28,29]. Several factors have been proposed to explain this behavior, including subtle cobalt stoichiometry [28] and sulfur vacancies [29]. However, a consistent understanding of how these experimental parameters influence material properties remains lacking.

Here, we investigate how cobalt composition influences the bulk properties of $Co_{1/3(1\pm\delta)}NbS_2$. Structural consistency across the target composition range was assessed by X-ray diffraction and Raman spectroscopy. Transport, magnetic susceptibility, and heat capacity measurements were then used to verify the composition dependence of the magnetic and electronic properties. The evolution of the Hall effect, including the suppression of the THE with composition, reflects a composition-driven magnetic transition in this compound. Specific heat measurements further reveal a clear signature of composition-dependent changes in the low-energy electronic contribution, reflected in the Sommerfeld coefficient. The experimentally observed M-point modulation vector and the corresponding triple-**Q** magnetic order are discussed within a microscopic spin-Hamiltonian framework, highlighting the role of higher-order exchange interactions in stabilising multi-**Q** magnetic order. These results indicate that $Co_{1/3(1\pm\delta)}NbS_2$ provides a model system with composition-tunable magnetic and electronic properties that remain largely decoupled from structural complexity. This work provides experimental insight into competing orders in intercalated TMDs controlled by fine-tuning of the intercalant concentration.

## Methods

Single crystals of $Co_{1/3(1\pm\delta)}NbS_2$ ($-4\% \leq \delta \leq 8\%$) were synthesized through a two-step process. Polycrystalline precursors were first prepared with solid-state reaction with high-purity elements (Cobalt (Alfa Aesar, 99.99%), Niobium (Alfa Aesar, 99.98%), and Sulfur (Sigma Aldrich, 99.99%)) at 900°C for 120 hours in evacuated quartz tubes (4 - 5 x $10^{-2}$ torr). To systematically investigate stoichiometry effects, the nominal cobalt concentration was varied from -4% to +8% relative to the stoichiometric value x = 1/3. In this study, we used this percentage deviation as our sample identifier; for instance, δ = 0% denotes the ideal $Co_{1/3}NbS_2$ composition, while δ = -2% and +4% denote samples with cobalt deficiency and excess phases during the solid-state reaction step. Single crystals were grown from the precursors using the chemical vapor transport (CVT) method. For a typical growth, 1.2 g of precursor and 0.2 g of iodine ($I_2$) were vacuum-sealed in a quartz tube (4 - 5 x $10^{-2}$ torr). After two weeks, with a temperature gradient from 950 °C to 850 °C, a typical centimeter-sized crystal could be obtained.

The structural properties of the synthesized samples were characterized by powder X-ray diffraction (Miniflex XRD; Rigaku) using Cu $K_\alpha$ radiation, and Rietveld refinement was performed with the FullProf Suite. Two complementary methods were used to verify the compositional homogeneity of the single crystals. Raman spectroscopy was performed with a 532 nm laser excitation (XPER RAM; Nanobase), along with energy-dispersive X-ray spectroscopy (QUANTAX; Bruker). Magnetic susceptibilities were measured using a SQUID magnetometer (MPMS3; Quantum Design). Electrical transport measurements, including resistivity and Hall effect, were performed in a superconducting magnet system (CFMS-9T; Cryogenic Ltd.) using a standard five-probe method. Specific heat was measured under high vacuum using the thermal relaxation method with a home-built sample puck (PPMS; Quantum Design).

## Results and Discussion

In this section, we address the composition dependence of transport, magnetic, and thermodynamic properties of $Co_{1/3(1\pm\delta)}NbS_2$. As discussed below, we found that the system's response to composition change not only modulates the magnetic properties inferred from the THE but also the electronic properties inferred from ordinary Hall coefficients and Sommerfeld coefficients. Additionally, we discuss the origin of the THE using a spin-Hamiltonian-based analysis, particularly for the triple-**Q** order with an M-point modulation vector, which has recently been identified as a key component of the THE in this system, while further contextualizing these findings through a comparative analysis with $Co_{1/3}TaS_2$ to elucidate the macroscopic transition pathways and the absence of electronic nematicity.

### A. Structural baseline and composition control

The structural characteristics of $Co_{1/3(1\pm\delta)}NbS_2$ single crystals are summarized in Fig.1. As illustrated in Fig. 1(a), Co atoms are intercalated within the Van der Waals gap of the $2H$-$NbS_2$ host, forming an AB-stacked triangular superlattice that gives rise to a $\sqrt{3}a_0 \times \sqrt{3}a_0$ superstructure with respect to the base $NbS_2$ lattice constant $a_0$. The resulting single crystals exhibit shiny, centimeter-sized hexagonal facets whose edges correspond to the crystallographic $a^*$-axis of $Co_{1/3}NbS_2$ crystal, which maps to the $a$-axis of the original $NbS_2$ host [Fig. 1(b)].

Evidence for long-range structural order of the cobalt superlattice is obtained from several experiments. An image from single-crystal X-ray diffraction (XRD) for the [H K 0] plane [Fig. 1(b)] shows superlattice reflections (with respect to the $2H$-$NbS_2$ structure) rotated by 30°, providing direct evidence for a periodic cobalt triangular network. Powder XRD patterns with Rietveld refinement further confirm the structural consistency of the intercalated phase [Fig. 1(c)]. We note that all structures were refined with the crystal structure to the space group $P6_322$ (No. 182), and the refinement results demonstrated reasonable agreement with the experimental data, as shown in Table 1. Raman spectroscopy [Fig. 1(d)] reveals nearly identical phonon features, including a characteristic mode around 150 - 200 $cm^{-1}$. To confirm the macroscopic spatial homogeneity of our crystals, Raman spectra were collected from 10 randomly selected spots per sample, yielding highly reproducible profiles. A closer inspection of the intercalation-induced mode around 150–200 $cm^{-1}$ reveals a gradual and systematic evolution of the peak center within our composition window (–4% ≤ δ ≤ +4%), before exhibiting a deviation at the highly over-intercalated regime (δ = +8%). The detailed results are provided in Appendix A. Together, these observations indicate that the basic structure remains preserved across the investigated composition window. The noticeable change appears in the $c$-axis lattice constant, which varies linearly with nominal cobalt concentration [Fig. 1(e)]. Furthermore, energy-dispersive X-ray (EDX) analysis demonstrates a systematic trend in which the actual chemical stoichiometry aligns well with the expected nominal variations [see Table 2]. Overall, $Co_{1/3(1\pm\delta)}NbS_2$ maintains structural consistency across the composition window investigated here. Therefore, the substantial changes in bulk properties discussed below are more naturally attributed to electronic and magnetic origin rather than structural type.

Regarding the over-stoichiometric compositions (δ > 0), a natural question arises concerning the specific microscopic accommodation of the excess Co ions. Although definitive identification of their exact location falls beyond the resolution of our macroscopic characterizations, structural considerations and prior studies on analogous compounds provide a plausible scenario [27]. Given that the ideal 1/3 stoichiometry corresponds ($Co_{1/3}NbS_2$) to a fully occupied $2c$ (or $2d$) Wyckoff site, it is most likely that any excess Co atoms are accommodated into the adjacent, previously vacant $2d$ (or $2c$) interstitial sites.

### B. Composition-dependent electronic and magnetic responses

Structural consistency is preserved across the studied composition range, while the electronic and magnetic properties of $Co_{1/3(1\pm\delta)}NbS_2$ exhibit a pronounced evolution with respect to subtle shifts in composition. The bulk properties fall naturally into three distinct trends. These three behaviors are highlighted in the figures using red, green, and blue color.

For the transport measurements, pristine single crystals were carefully selected and mechanically shaped into rectangular bar geometries using a razor blade. Typical dimensions were 5 - 7 mm in lateral size, and thicknesses ranged from 10 to 30 μm. First, the distinct evolution across the cobalt compositions can be traced through the longitudinal resistivity ($\rho_{xx}$) and magnetic susceptibility ($M/H$). Fig. 2(a) shows that the temperature dependence of $\rho_{xx}$ reveals that all compositions exhibit nearly identical behavior above $T_N$, a feature clearly detectable in the temperature derivative of resistivity, $d\rho_{xx}/dT$. However, below the magnetic ordering temperature, qualitative differences emerge across the compositions. The most dramatic changes occur near the ideal stoichiometry (δ = 0% and +2%), where the onset of magnetic order leads to a pronounced drop in resistivity. This trend is further corroborated by the magnetic susceptibility data presented in Fig. 2(b). In the composition range where the THE is observed (δ = -4% to +2%), a characteristic peak-like signature appears at $T_N$, accompanied by a noticeable difference between zero-field-cooled (ZFC) and field-cooled (FC) measurements. Although the detailed behavior differs among compositions, this kind of difference in ZFC - FC measurements occurs in compositions with THE. We note that the two distinct peaks observed in the magnetic susceptibility for δ = +2% do not signify two separate magnetic phase transitions, but rather originate from the interplay between a low-temperature weak moment reversal (discussed below) and a field-induced critical response near $T_N$. On the contrary, for the case of δ = 4% and 8%, where the THE is absent, the resistivity anomaly is significantly smeared, and the magnetic susceptibility shows only a marginal difference between ZFC and FC measurements. It is noted that for δ = 8% samples, the temperature of the resistivity anomaly deviates from the position of the cusp in magnetic susceptibility, potentially suggesting a gradual decoupling of itinerant electrons and localized moments.

Before discussing the composition dependence in the Hall effect, we briefly rationalize the observed Hall resistivity ($\rho_{yx}$) to elucidate its topological origin and distinguish it from potential anomalous Hall effect (AHE), defined here in a narrow sense as a contribution strictly proportional to the bulk magnetization ($M$). Crucially, the scaling between the spontaneous Hall effect and the bulk magnetization reveals two major quantitative contradictions for a standard AHE scenario. First, while the observed Hall angle reaches a massive scale of $\theta_H \sim O(10^{-2})$, the weak ferromagnetic moment (WFM) is extraordinarily small, only $O(10^{-3})\mu_B/Co$ [see Fig. 3(b) and 3(c)]; attributing this Hall signal solely to AHE would require an unphysically large anomalous Hall coefficient. Second, upon field-cooled training, the WFM in specific compositions (δ = 0% and +2%) exhibits a clear sign reversal upon cooling [see Fig. 3(c)]. In contrast, the spontaneous Hall effect strictly maintains its sign without any corresponding inversion. Furthermore, the nearly monotonic field dependence of the $M(H)$ curves reported previously [20,29] effectively rules out multi-component AHE artifacts. Taken together, along with the indirect signature from recent polarized neutron diffraction [21], these results robustly demonstrate that the spontaneous Hall signal is an intrinsic property originating from a non-coplanar chiral spin texture.

Changing the cobalt composition strongly modifies the electronic and magnetic behavior of $Co_{1/3(1\pm\delta)}NbS_2$, as reflected in the Hall measurements [Fig. 3]. In the composition range where a topological Hall effect is active (δ = -4% to +2%), a pronounced hysteresis loop is observed at T = 27 K [Fig. 3(a)]. Also, a clear difference emerges in the temperature dependence of the Hall response. While the -4% and -2% Co samples show a relatively featureless evolution from $T_N$, the samples with δ = 0% and +2% show a broad peak-like feature centered around $T \approx 25\ K$ for the spontaneous Hall effect [Fig. 3(b)]. This feature is consistently seen in the ordinary Hall coefficient $R_H$ [Fig. 4], which also shows a notable enhancement at the same temperature. This similar evolution of the spontaneous Hall effect $[\rho_{yx}(T; H = 0)]$ and $R_H(T)$ indicates that another modification of the electronic response occurs below the ordering temperature, suggesting that transport properties are not solely governed by the onset of magnetic ordering.

A closer inspection reveals that the longitudinal ($\rho_{xx}$) and Hall ($\rho_{yx}$) resistivity exhibits distinct behavior for the δ = 0% and +2% samples. These two compositions exhibit a sharp drop in resistivity upon magnetic order. Given the robust structural consistency of these compounds and the onset of resistivity drop concurrent with $T_N$, this behavior is naturally attributed to a sharp reduction in the spin-fluctuation scattering rate, $\tau_{\text{spin}}^{-1}$ within the framework of Matthiessen's rule, Eq. (1) [30]. To qualitatively isolate the magnetic contribution to the transport properties, we model the temperature dependence of $\rho_{xx}$ using a phenomenological formula adopted from previous study [31]. Based on the spirit of Matthiessen's rule, the total resistivity in the paramagnetic regime ($T$ = 50 – 300 K) is treated as an additive combination, expressed as Eq. (2):

$$\tau_{\text{sum}}^{-1} = \tau_{\text{impurity}}^{-1} + \tau_{\text{phonon}}^{-1} + \tau_{\text{spin}}^{-1} + (\dots) \tag{1}$$

$$\rho_{xx}(T) = \rho_0 + \rho_1 \left(\frac{T}{\Theta_D}\right)^n \int_0^{\frac{\Theta_D}{T}} \frac{x^n e^x}{(e^x-1)^2} dx + \rho_2 e^{-\frac{T_0}{T}} \tag{2}$$

The fitting curve is illustrated as an orange dashed line in Fig. 2(a) with the fitting parameter provided in Table 3 for each composition. This framework provides a consistent framework for evaluating scattering mechanisms across different temperature regimes. In the high-temperature regime above $T_N$, the adopted formula successfully captures the experimental behavior, demonstrating that the paramagnetic transport is reasonably described by two main scattering mechanisms. Below the magnetic ordering temperature ($T_N$), however, the experimental data systematically deviate from this fitting baseline, reflecting the additional reduction in the scattering rate driven by the spin ordering. Still, the emergence of the additional feature at $T \approx 25\ K$ in the Hall effect, distinct from $T_N$ and appearing only in a narrow composition window, points toward a more complex origin rather than a simple reduction in magnetic scattering. One could speculate that such behavior implies the presence of an additional electronic contribution, whether induced by magnetic ordering or by some unknown cause. This potential electronic modification scenario is further supported by the distinct Sommerfeld coefficient ($\gamma$) observed in the composition dependence of the heat capacity, as discussed below.

Fig. 5 summarizes the transverse and longitudinal conductivity in the temperature-composition phase diagram, with distinct dependence between $\sigma_{xx}$ and $\sigma_{yx}$ clearly observed. Combined with the additional cusp observed in the spontaneous Hall effect and the ordinary Hall coefficient, these results suggest that composition tuning not only influences the magnetic interaction but also the electronic structure, as discussed in terms of heat capacity below.

We note that a previous report on a transport study on $Co_{1/3(1\pm\delta)}NbS_2$ reported a sign reversal in the field-tunable topological Hall effect for certain compositions with a high residual resistivity ratio (RRR) [28]. Although this sign reversal is not observed in our current transport measurements, we believe that the intrinsic magnetic behavior remains consistent. Previous reports have established that the hysteresis loop of the WFM in this system is synchronous with the switching field of the Hall resistivity, indicating that the WFM is intimately coupled with the chirality of the spin structure, which dictates the sign of the Hall response. The temperature dependence of this WFM shows a clear composition dependence; the δ = -4% and -2% samples maintain a constant sign, whereas the δ = 0% and +2% samples (which show the high RRR) exhibit a distinct sign reversal during cooling [see Fig. 3(c)]. The absence of the actual Hall loop reversal in our transport measurements can be understood in terms of the behavior of the coercive field ($H_c$). A larger $H_c$ in our current crystals poses a practical challenge in capturing the full switching within our experimental field limits (±9 T). Therefore, these observations collectively suggest a consistent underlying mechanism between the two studies, where the lack of a visible sign reversal in our case is primarily driven by the large coercive field $H_c$ rather than implying a fundamentally distinct mechanism.

Regarding the large coercive field in $Co_{1/3(1\pm\delta)}NbS_2$, we note that this magnetic hardness can be effectively contextualized by establishing a comparative framework with its isostructural sister compound $Co_{1/3}TaS_2$, particularly through the distinct scale of its spontaneous moment (or WFM) compared to $Co_{1/3}TaS_2$ [18,20,25]. As consistently demonstrated in previous literature, the reversal of the Hall effect is tightly coupled to that of the WFM. While the field-driven reversal of the spin chirality is associated with the WFM, the magnitude of this spontaneous moment in $Co_{1/3}NbS_2$ $[O(10^{-3})\mu_B/Co]$ is an order of magnitude smaller than that in $Co_{1/3}TaS_2$ $[O(10^{-2})\mu_B/Co]$, where $H_c$ reaches around 5 T for 5 K [18]. Since the external field couples with this WFM to provide the Zeeman energy required to flip the magnetic domains, a significantly higher magnetic field is necessary to overcome the energy barrier when the spontaneous moment is inherently small. This difference in Zeeman coupling strength provides a plausible explanation for the much higher coercive fields observed in $Co_{1/3(1\pm\delta)}NbS_2$ than in $Co_{1/3}TaS_2$ at a qualitative level. In this context, the variation of $H_c$ at 27 $K$ [Fig. 3(a)] reflects a sensitivity of coercive field near the transition temperature, given that $H_c$ increases abruptly below $T_N$ across all compositions. Specifically, the larger loops for δ = 0% and +2% arise because their slightly higher $H_c$ places them thermodynamically deeper within the ordered phase at 27 K compared to δ = -4% and -2%. We note that for all compositions (-4% ≤ δ ≤ +2%), the coercive field rapidly exceeds the 9 T limit of our superconducting magnet upon further cooling, a behavior consistent with previous reports [13,20].

## C. Heat capacity measurement

To reveal the thermodynamic nature of each composition, we measured the heat capacity for the δ = -2%, +2%, and +4% Co samples [Fig. 6(a)]. Consistent with the resistivity anomalies, the δ = +2% sample exhibits the most prominent peak at the magnetic ordering temperature, $T_N$. To isolate the magnetic contribution to the heat capacity ($C_{\mathrm{mag}}$), the non-magnetic background from the phonon contribution was subtracted off using a combination of one Debye and two Einstein models, based on phonon calculation using density functional theory [32]. The electronic background was subsequently removed using the Sommerfeld coefficient ($\gamma$) described by the method discussed below. The magnetic entropy ($S_{\mathrm{mag}}$), extracted by subtracting the lattice and electronic contributions, was compared across the compositions. We found that the δ = -2% and +2% samples approached the theoretical value of a $S = 3/2$ system. In contrast, the +4% Co samples showed a slight deficiency in magnetic entropy. This reduction of magnetic entropy coincides with a broader and smaller magnetic peak in $C_p/T$, as well as slight splitting observed in ZFC/FC measurement of magnetic susceptibility. These results suggest that the spins may not fully participate in the long-range magnetic ordering at $T_N$, as it exhibits a freezing-like behavior.

Lattice ($\beta$) and electronic ($\gamma$) contributions to specific heat are extracted from low-temperature behavior ($C/T - T^2$) [Fig. 6(b)]. The Sommerfeld coefficient ($\gamma$) shows a clear discrepancy compared to the lattice part, which remains nearly identical. Since the ordinary Hall coefficient ($R_H$) also varies distinctly, we examined the relationship between the $R_H^{-1}$ and the $\gamma$. The two quantities exhibit a linear scaling [Fig. 6(b)]. This relationship aligns with a single-band picture, as the electronic transport remains consistently hole-dominant, with no sign change in the Hall effect across the studied temperature and composition range. Taken together, these observations indicate that the cobalt concentration systematically tunes the electronic response, reflecting a coherent evolution of the electronic state that extends beyond a simple dilute-impurity picture.

The correlated evolution of these bulk properties motivates us to consider a composition-driven reorganization of the electronic structure in this system. Although the consistency in crystal structure and magnetic transition temperatures across the studied range makes a drastic reconstruction of the electronic structure less probable, such macroscopic robustness does not preclude minute, local modifications in the low-energy electronic landscape. A previous angle-resolved photoemission spectroscopy (ARPES) study on $Co_{1/3}NbS_2$ and $Co_{1/3}NbS_{2-\delta}$ [29] has already demonstrated that subtle changes in stoichiometry could lead to a rigid shift of the Fermi level by approximately 50 meV. A compelling precedent is also found in the sister compound, $Co_xTaS_2$ (x ≈ 1/3), where a characteristic band reconstruction specifically at the M-point was observed within a specific composition window [33].

While these findings alone do not provide direct evidence for a Fermi surface reconstruction, the systematic scaling of $\gamma$ and $R_H^{-1}$ and the previous ARPES results strongly point toward a composition-selective evolution of the low-energy electronic structure. This picture is further supported by the transport response, where a clear transport anomaly in the Hall effect appears near $T \approx 25\ K$ only within a limited composition range (δ = 0% and 2%). The fact that this feature is not observed across all compositions suggests that it is not a simple consequence of magnetic ordering. Collectively, our results suggest a possible reconstruction of the low-energy electronic structure over a narrow composition range in $Co_{1/3(1\pm\delta)}NbS_2$.

To preclude any potential misinterpretation, we clarify that this change in electronic state is decoupled from the presence of the topological Hall effect itself. The concurrent manifestation of long-range antiferromagnetic ordering and a non-zero spontaneous topological Hall effect across the window from δ = -4% to +2% strongly suggests that the underlying magnetic configuration is a non-coplanar spin structure. This independence is also evidenced by the δ = -2% sample; despite exhibiting a larger Sommerfeld coefficient (γ) than the suppressed value observed in the δ = +2% composition, it concurrently maintains a remarkably robust topological Hall response. Therefore, while the non-coplanar chiral spin texture serves as a universal ground-state baseline across a broader window, the local γ minimum implies an independent, composition-selective tuning of the low-energy electronic structure.

### D. Modulation vector and role of higher-order interaction for multi-Q structure

The THE in $Co_{1/3(1\pm\delta)}NbS_2$ disappears without significant structural modification. While the microscopic mechanism underlying this behavior remains unresolved, the presence of THE is closely linked to a magnetic structure characterized by triple-**Q** order with an M-point modulation vector, **Q** = (1/2, 0, 0) [13,21]. This ordering is distinct from the canonical 120° spin structure observed in triangular-lattice magnets [34,35]. Understanding the origin of this modulation vector requires a microscopic spin-Hamiltonian description. Theoretical work suggests that higher-order exchange interactions sensitive to the Fermi surface, such as biquadratic terms, play an important role in stabilizing multi-**Q** spin order [36–38]. The disappearance of THE could therefore reflect a reduction or modification of these higher-order interactions caused by a composition-induced shift of the Fermi surface. Such a scenario explains the fragility of the topological phase without requiring a substantial change in the primary bilinear exchange interactions, which are expected to remain relatively robust to small variations in stoichiometry. This singular sensitivity near the ideal 1/3 stoichiometry is a recurring theme in intercalated transition-metal dichalcogenides; for instance, $Fe_{1/3(1\pm\delta)}NbS_2$ and $Co_{1/3(1\pm\delta)}TaS_2$ also exhibit a sharp electronic and magnetic phase boundary where the ground state is exceptionally susceptible to minor compositional deviations [16,17,25]. Such precedents suggest that the 1/3-filling marks a critical regime where the interplay between structural symmetry and electronic states is most delicate. Below, we examine the spin Hamiltonian required to stabilize the M-point modulation and analyze how the strength of the biquadratic interaction controls the transition between the single-**Q** and triple-**Q** magnetic states.

The triangular lattice system has long been regarded as a canonical platform for exploring geometric frustration, so-called 120° structure (corresponds to **Q** = (1/3, 1/3, 0) or K-point in reciprocal space) is typically stabilized by antiferromagnetic nearest-neighbor exchange interactions ($J_1 > 0$) [34,35,39]. However, recent reports on $Co_{1/3}NbS_2$ and $Co_{1/3}TaS_2$ have identified a magnetic modulation vector of **Q** = (1/2, 0, 0) (M-point), which deviates from the canonical 120° structure [21]. To understand the physical origin of the stabilization of this M-point order over the K-point order, we investigate the magnetic Hamiltonian using the Luttinger-Tisza method [40]. While previous theoretical studies have primarily discussed the magnetic modulation vector in the context of electronic Hamiltonians based on tight-binding models [36–38], we focus on identifying which modulation vectors are stabilized within a spin Hamiltonian framework. The spirit of the Luttinger-Tisza method lies in analyzing the energy minimum of the Fourier-transformed exchange interaction, $J(\boldsymbol{q})$, to determine the stability of the modulation vector for a given spin Hamiltonian.

$$H = \sum_{n,\boldsymbol{\delta}} J_{\boldsymbol{\delta}}(\boldsymbol{S}_n \cdot \boldsymbol{S}_{n+\boldsymbol{\delta}}) \rightarrow J(\boldsymbol{q}) = \sum_{\boldsymbol{\delta}} J_{\boldsymbol{\delta}} e^{i\boldsymbol{q}\cdot\boldsymbol{\delta}} \quad (3)$$

where *n* and $\boldsymbol{\delta}$ denote site indices and displacement, $\boldsymbol{S}_n$ and $J_{\boldsymbol{\delta}}$ denotes spin at site *n* and exchange interaction for given displacement $\boldsymbol{\delta}$.

Crucially, recent studies on the isostructural system $Co_{1/3}TaS_2$ have revealed that the interlayer exchange interaction ($J_{c1}$) can be comparable to or even exceed the intra-layer nearest-neighbor exchange ($J_1$) [41]. Given the structural similarity, it is reasonable to assume that strong interlayer coupling is present in $Co_{1/3}NbS_2$. This provides a physical basis for investigating the influence of $J_{c1}$ and $J_{c2}$ on the magnetic ground state [see Fig. 7(a) for geometrical details of exchange interaction]. As illustrated in Fig. 7(c), the tuning of interlayer parameters profoundly reshapes the phase diagram from the 2D $J_1 - J_2 - J_3$ model [see Fig. 7(b) for the mapping of the colormap with a modulation vector]. Specifically, as the antiferromagnetic $J_{c1}$ increases, the canonical 120° order, **Q** = (1/3, 1/3, 0), is rapidly pushed out outside of the focus region of parameter space. Moreover, the introduction of a non-negligible $J_{c2}$, consistent with its reported role in $Co_{1/3}TaS_2$, explains the stabilization of the **Q** = (1/2, 0, 0) order within the explored parameter space.

Along with the stabilization of the magnetic modulation vector, the emergence of a multi-**Q** (triple-**Q**) structure over an energetically degenerate single-**Q** structure remains a critical question. Recent theoretical works [36–38] provide an effective Hamiltonian-based approach to this question in terms of a bilinear-biquadratic Hamiltonian with positive biquadratic interaction.

$$H = \sum_n \sum_\delta J_\delta(\boldsymbol{S}_n \cdot \boldsymbol{S}_{n+\delta}) + \sum_n \sum_\delta K_\delta(\boldsymbol{S}_n \cdot \boldsymbol{S}_{n+\delta})^2 \quad (4)$$

where $n$ and $\boldsymbol{\delta}$ denote site indices and displacement, $J_\delta$ and $K_\delta$ correspond to the exchange and biquadratic interaction for given displacement $\boldsymbol{\delta}$. These higher-order interactions (e.g., biquadratic interaction) are theoretically predicted to contribute non-negligibly when Fermi surface nesting conditions are satisfied [36–38]. In this formalism, a pure bilinear exchange Hamiltonian yields single-**Q** and multi-**Q** structures that degenerate in energy.

However, the introduction of higher-order interactions beyond bilinear exchange, specifically a positive biquadratic interaction (*K*), can lift this degeneracy and lead to a triple-**Q** magnetic ground state. Previous studies on $Co_{1/3}TaS_2$ suggest that the transition pathway toward a multi-**Q** state depends on the form of the Hamiltonian, particularly on the strength of the biquadratic interaction [23]. $Co_{1/3}TaS_2$ exhibits a temperature-driven paramagnetic - single-**Q** - triple-**Q** transition [23]. In $Co_{1/3(1\pm\delta)}NbS_2$, we observe a direct transition from the paramagnetic state to the triple-**Q** state without an intermediate phase. The presence of a two-step transition is controlled by the strength of a positive nearest-neighbor biquadratic interaction ($K$), as indicated by previous Landau-Lifshitz-Gilbert (LLG) simulations, effectively providing a lower limit of this interaction in $Co_{1/3(1\pm\delta)}NbS_2$ [23].

To examine this possibility, LLG simulations were carried out using the *Su(n)ny* package [42] implemented in *Julia* programming language [43]. Since the precise exchange constants are not yet known, we adopted a minimal $J_1 - J_2 - J_{c1}$ Hamiltonian with finite biquadratic interaction ($K$) for nearest neighbor as a representative framework for the system. The parameters were chosen as $J_1 = 1.85$ meV, $J_{c1} = 0.5\,J_1$, and $J_2 = 0.2\,J_1$, which were specifically chosen to stabilize the M-point modulation vector and reproduce the experimental ordering temperature ($T_N \approx 30\ K$). Calculations used a 20 × 20 × 10 supercell. Each quantity averaged over 5000 steps for each temperature after 2000 steps of thermalization steps. As $K$ increases, the phase boundary between single-**Q** and triple-**Q** order shifts [see Fig. 8]. With increasing $K$, the temperature window of the single-**Q** phase narrows and eventually yields a direct paramagnetic - triple-**Q** transition. The exact position of this boundary depends on the choice of exchange Hamiltonian; the minimal model reproduces the overall trend. The simulations show that long-range bilinear exchange combined with higher-order biquadratic interactions can stabilize the M-point modulation vector with triple-**Q** order. Given that recent advances have enabled the rigorous calculation of biquadratic interactions from density functional theory, further investigations would be valuable in elucidating the microscopic origin of multi-**Q** order in this system [44].

We emphasize that this effective spin Hamiltonian approach is intended as a conceptual framework to illustrate the fundamental mechanism behind the direct stabilization of the triple-**Q** state, rather than proposing a strict one-to-one mapping to the experimental cobalt composition δ). Additionally, we reserve judgment on the exact magnetic ground state of the over-doped regime (δ ≥ +4%). We do not assert that it necessarily corresponds to an M-point single-**Q** order, as this remains to be verified by future studies.

Furthermore, our choice to employ a uniform minimal model—without explicit terms for Co vacancies or interstitial defects—is physically justified by the statistically dilute nature of the off-stoichiometry within our primary focus region (-4% ≤ δ ≤ +4%), where the key magnetic and resistivity anomalies coincide. The sparcity of these defects can be described as follows in the context of the macroscopic magnetic structure: a two-dimensional domain of 10 × 10 Co sites encompasses 25 magnetic unit cells of the M-point triple-**Q** order. At a maximum stoichiometry deviation of 4%, only 4 defects are introduced within this entire domain. Consequently, even under a scenario of maximal defect dispersion, 21 out of 25 magnetic unit cells remain completely pristine and unperturbed by any local impurity.

Furthemore, the large nearest-neighbor distance between Co atoms (*d* ~ 5.7 Å) effectively suppresses direct exchange mechanisms, such as super-exchange interaction. Given that the magnetic coupling in this system is therefore predominantly governed by long-range RKKY interactions, the localized effects of these statistically sparse impurities are expected to be averaged out across the pristine majority of the lattice. Therefore, these defects act as minor local perturbations rather than fundamental drivers of the macroscopic magnetic interaction, allowing our minimal Hamiltonian—focusing on the essential exchange parameters (*J* and *K*)—to effectively describe the primary physics of the system.

### E. Thermal fluctuation, magnetic anisotropy and symmetry of the triple-Q phase

To contextualize the distinct phase evolution observed in $Co_{1/3(1\pm\delta)}NbS_2$, it is instructive to establish a comparative framework with its isostructural sister compound $Co_{1/3}TaS_2$ by examining the interplay between higher-order interactions and exchange and single-ion anisotropy. Theoretically, the emergence of an intermediate single-**Q** phase is rooted in a competition between internal energy and thermal entropy. While the positive biquadratic interaction (*K*) successfully lifts the classical degeneracy to establish the non-coplanar triple-**Q** structure as the absolute ground state at zero temperature, thermal fluctuations right below $T_N$ can entropically favor a collinear magnetic configuration. This behavior is physically consistent with the "order-by-thermal-disorder" mechanism [45], where the collinear single-**Q** phase becomes thermodynamically stable over a finite temperature window when the biquadratic term is only moderate. Experimentally, however, such an intermediate phase does not stabilize in our $Co_{1/3(1\pm\delta)}NbS_2$ samples across the under-intercalated and stoichiometric regimes (δ ≤ +2%), which instead exhibit a single, direct transition into the triple-**Q** state at $T_N$, as evidenced by the simultaneous emergence of the spontaneous Hall effect alongside distinct anomalies in both resistivity and magnetic susceptibility. This clear distinction implies that the biquadratic interaction in $Co_{1/3}NbS_2$ is significantly stronger than that in its sister compound $Co_{1/3}TaS_2$ [23], being sufficiently robust to completely bypass the fluctuation-dominated intermediate single-**Q** regime, thereby setting a lower bound for this interaction in our system.

As a qualitative picture, this difference in the effective $K$ between these sister compounds could be taken as evidence for the validity of the isotropic RKKY framework. For $Co_{1/3}TaS_2$, the hosting Ta 5$d$ layers would possess a stronger spin-orbit coupling (SOC) compared to the Nb 4$d$ layers. Through hybridization with intercalated Co ions, this strong SOC in the host layers can introduce local anisotropy at the magnetic sites, potentially perturbing the purely isotropic bilinear-biquadratic RKKY interaction profiles. In contrast, for $Co_{1/3}NbS_2$, the relatively weak SOC of the Nb layers suggests that the system remains closer to an isotropic regime. This weaker perturbation is expected to preserve a larger effective $K$ parameter, which can overcome thermal entropy right below $T_N$ and drive a direct transition into the triple-**Q** state.

In line with this qualitative picture, incorporating magnetocrystalline anisotropy through both single-ion and exchange anisotropy terms helps illustrate the symmetry of the stabilized triple-**Q** state, as exemplified by previous optical studies on $Co_{1/3}TaS_2$ [26]. Specifically, when anisotropic terms such as bond-dependent exchange and single-ion anisotropy are considered, they break the equal intensity among the constituent $\mathbf{Q}_1$, $\mathbf{Q}_2$, and $\mathbf{Q}_3$ vectors [46]. These anisotropic interactions stabilize a distorted triple-**Q** phase as the ground magnetic structure, which is characterized by unequal amplitudes of the corresponding Fourier components ($|\boldsymbol{S}_{\boldsymbol{Q}_1}|$, $|\boldsymbol{S}_{\boldsymbol{Q}_2}|$ and $|\boldsymbol{S}_{\boldsymbol{Q}_3}|$), a scenario suggested for the results of $Co_{1/3}TaS_2$ [26]. Conversely, when such anisotropic interactions are absent, the system settles into a symmetric triple-**Q** phase where the corresponding Fourier components maintain equal amplitudes, as shown in our current simulation results [see Fig. 8(c)], which we consider to be the case for $Co_{1/3}NbS_2$.

Furthermore, this difference in magnetic structure provides a consistent framework for considering the reported trends regarding electronic nematicity in these materials. In $Co_{1/3}TaS_2$, the single-**Q** or distorted triple-**Q** phases both break the lattice rotational symmetry (specifically $C_3$ rotation symmetry), which can serve as a compatible background to host electronic nematicity [11,26,47]. Conversely, for $Co_{1/3}NbS_2$, the direct transition into a symmetric triple-**Q** state suggests that the $C_3$ rotation symmetry remains macroscopically preserved from the onset of ordering. Under this scenario, the symmetry-broken magnetic structure, required to sustain electronic nematicity, is not established, offering a plausible explanation for why nematic behaviors appear to be suppressed in $Co_{1/3}NbS_2$. Ultimately, this comparative discussion underscores that while the biquadratic interaction dictates the macroscopic transition pathway (paramagnet → triple-**Q** or paramagnet → single-**Q** → triple-**Q**), the local magnetic anisotropy fine-tunes the internal symmetry of the multi-**Q** state (distorted triple-**Q** or symmetric triple-**Q**), providing a comprehensive perspective on the diverse magnetic ground states in intercalated transition metal dichalcogenides.

## Conclusion

In summary, our systematic investigation reveals that the bulk properties of $Co_{1/3(1\pm\delta)}NbS_2$, including THE, are remarkably sensitive to the precise cobalt stoichiometry. We identified a critical composition window near the ideal stoichiometry, δ = 0 - 2%, where a significant enhancement in longitudinal conductivity occurred before the suppression of THE. Within this same narrow range, the ordinary Hall effect exhibits a distinct enhancement where $R_H^{-1}$ scales linearly with the Sommerfeld coefficient ($\gamma$). This correlation strongly suggests that cobalt concentration serves as a direct knob for tuning the low-energy electronic structure in this system.

An intriguing finding of this work is that while the THE vanishes in the δ = +4% sample, its other bulk properties remain qualitatively similar to those of the Hall-active δ = -2% sample. This observation raises fundamental questions regarding the stability of the **Q** = (1/2, 0, 0) magnetic modulation vector across the phase boundary. Unlike other intercalated transition metal dichalcogenides (TMDs), where composition sensitivity is often driven by a shift in the magnetic modulation vector [17,25], our results suggest that $Co_{1/3(1\pm\delta)}NbS_2$ may represent a unique case where such sensitivity occurs through a subtle reconstruction of the electronic landscape without a change in the primary magnetic symmetry. Further studies utilizing transport probes highly sensitive to the Fermi level, such as thermoelectric or Nernst effect measurements, will be essential to fully elucidate this microscopic evolution of the electronic structure and its complex interplay with the underlying magnetic order.

## Acknowledgement

We acknowledge Pyeongjae Park, Scott A. Crooker, Taekoo Oh, Jeonghun Kang, Kwang-Tak Kim, Beom Hyun Kim, and Hyowon Park for their helpful discussions. This work was supported by the Samsung Science & Technology Foundation (Grant No. SSTF-BA2101-05) & the Core Center Program (2021R1A6C101B418) by the Ministry of Education. One of the authors (J.-G.P.) is partly funded by the Leading Researcher Program of the National Research Foundation of Korea (Grant No. RS-2020-NR049405).

## Appendix A. Spatial homogeneity and compositional dependence of Raman spectra

To robustly evaluate the macroscopic spatial homogeneity of the $Co_{1/3(1\pm\delta)}NbS_2$ single crystals and to rule out potential artifacts arising from structural phase separation, we measured Raman spectra at multiple random spots. For each composition, spectra were collected at 10 randomly selected locations directly on the surface of the identical crystal utilized for the transport and Hall measurements. As shown in Fig. A1(a), the Raman profiles obtained from these disparate spots exhibit a highly reproducible line shape, verifying the structural uniformity across the macroscopic scale of the transport samples.

A closer quantitative inspection focuses on the intercalation-induced phonon mode in the frequency range of 150 – 200 $cm^{-1}$, which reflects the formation of the intercalated Co superlattice. With decreasing Co concentration in the under-intercalated regime, the datasets reveal a systematic red-shift of the peak center ($\omega_0$) accompanied by a slight broadening of the linewidth ($\Gamma$), as summarized in Fig. A1(b). While this evolution leaves room for multiple interpretations of its detailed microscopic origin, the monotonic trend demonstrates that variation in Co concentration systematically alters the structural state of the system at the lattice vibration level. Notably, within our composition window ($-4\% \leq \delta \leq +4\%$) where magnetic and resistivity anomaly temperatures coincide, both $\omega_0$ and $\Gamma$ exhibit a nearly monotonic dependence on the nominal Co concentration.

## Reference

[1] L. Onsager, Crystal Statistics. I. A Two-Dimensional Model with an Order-Disorder Transition, Phys. Rev. **65**, 117 (1944).

[2] J. M. Kosterlitz and D. J. Thouless, Ordering, metastability and phase transitions in two-dimensional systems, J. Phys. C Solid State Phys. **6**, 1181 (1973).

[3] V.L.Berezinskii, *Destruction of Long-Range Order in One-Dimensional and Two-Dimensional Systems Possessing a Continuous Symmetry Group. II. Quantum Systems,* Sov.Phys.JETP **32**, 493 (1971).

[4] N. D. Mermin and H. Wagner, Absence of Ferromagnetism or Antiferromagnetism in One- or Two-Dimensional Isotropic Heisenberg Models, Phys. Rev. Lett. **17**, 1133 (1966).

[5] J.-U. Lee, S. Lee, J. H. Ryoo, S. Kang, T. Y. Kim, P. Kim, C.-H. Park, J.-G. Park, and H. Cheong, Ising-Type Magnetic Ordering in Atomically Thin FePS3, Nano Lett. **16**, 7433 (2016).

[6] J.-G. Park, Opportunities and challenges of 2D magnetic van der Waals materials: magnetic graphene?, J. Phys. Condens. Matter **28**, 301001 (2016).

[7] J.-G. Park, K. Zhang, H. Cheong, J. H. Kim, C. A. Belvin, D. Hsieh, H. Ning, and N. Gedik, 2D van der Waals magnets: From fundamental physics to applications, Rev. Mod. Phys. **98**, 025003 (2026).

[8] C.-T. Kuo et al., Exfoliation and Raman Spectroscopic Fingerprint of Few-Layer NiPS3 Van der Waals Crystals, Sci. Rep. **6**, 20904 (2016).

[9] S. S. Parkin and R. H. Friend, 3d transition-metal intercalates of the niobium and tantalum dichalcogenides ii. Transport properties, Philos. Mag. B **41**, 95 (1980).

[10] S. Yang, T. Zhang, and C. Jiang, van der Waals Magnets: Material Family, Detection and Modulation of Magnetism, and Perspective in Spintronics, Adv. Sci. **8**, 1 (2021).

[11] Z. Feng, W. Lu, T. Lu, F. Liu, J. R. Sheeran, M. Ye, J. Xia, T. Kurumaji, and L. Ye, Nonvolatile Nematic Order Manipulated by Strain and Magnetic Field in a Layered Antiferromagnet, 1 (2025): arXiv:2507.05486 (2025)).

[12] J. Kim, K.-X. Zhang, P. Park, W. Cho, H. Kim, H.-J. Noh, and J.-G. Park, Electrical control of topological 3Q state in intercalated van der Waals antiferromagnet Cox-TaS2, Nat. Commun. **16**, 8943 (2025).

[13] N. D. Khanh et al., Gapped nodal planes and large topological Nernst effect in the chiral lattice antiferromagnet CoNb3S6, Nat. Commun. **16**, 2654 (2025).

[14] P. Gu, Y. Peng, S. Yang, H. Wang, S. Ye, H. Wang, Y. Li, T. Xia, J. Yang, and Y. Ye, Probing the anomalous Hall transport and magnetic reversal of quasi-two-dimensional antiferromagnet Co1/3NbS2, Nat. Commun. **16**, 4465 (2025).

[15] H. H. Hill, Early actinides: the periodic system's f electron transition metal series, Nucl. Met., Met. Soc. AIME **17**, 2 (1970).

[16] S. Wu et al., Highly Tunable Magnetic Phases in Transition-Metal Dichalcogenide Fe1/3+δNbS2, Phys. Rev. X **12**, 021003 (2022).

[17] S. Wu et al., Discovery of Charge Order in the Transition Metal Dichalcogenide FexNbS2, Phys. Rev. Lett. **131**, 186701 (2023).

[18] P. Park, Y.-G. Kang, J. Kim, K. H. Lee, H.-J. Noh, M. J. Han, and J.-G. Park, Field-tunable toroidal moment and anomalous Hall effect in noncollinear antiferromagnetic Weyl semimetal Co1/3TaS2, Npj Quantum Mater. **7**, 42 (2022).

[19] Y. An et al., Bulk properties of the chiral metallic triangular antiferromagnets Ni1/3NbS2 and Ni1/3TaS2, Phys. Rev. B **108**, 054418 (2023).

[20] N. J. Ghimire, A. S. Botana, J. S. Jiang, J. Zhang, Y. S. Chen, and J. F. Mitchell, Large anomalous Hall effect in the chiral-lattice antiferromagnet CoNb3S6, Nat. Commun. **9**, 7 (2018).

[21] H. Takagi et al., Spontaneous topological Hall effect induced by non-coplanar antiferromagnetic order in intercalated van der Waals materials, Nat. Phys. **19**, 961 (2023).

[22] E. Kirstein, H. Park, I. Martin, J. F. Mitchell, N. J. Ghimire, and S. A. Crooker, Topological Magneto-optics in the Noncoplanar Antiferromagnet Co1/3NbS2, Phys. Rev. Lett. **135**, 196702 (2025).

[23] P. Park et al., Tetrahedral triple-Q magnetic ordering and large spontaneous Hall conductivity in the metallic triangular antiferromagnet Co1/3TaS2, Nat. Commun. **14**, 8346 (2023).

[24] A. Little et al., Three-state nematicity in the triangular lattice antiferromagnet Fe1/3NbS2, Nat. Mater. **19**, 1062 (2020).

[25] P. Park, W. Cho, C. Kim, Y. An, M. Avdeev, K. Iida, R. Kajimoto, and J.-G. Park, Composition dependence of bulk properties in the Co-intercalated transition metal dichalcogenide Co1/3TaS2, Phys. Rev. B **109**, L060403 (2024).

[26] E. Kirstein, P. Park, W. Cho, C. D. Batista, J. Park, and S. A. Crooker, Tunable chiral and nematic states in the triple-Q antiferromagnet Co1/3TaS2, Nature Communications **17**, 2212 (2026).

[27] H.-J. Noh, E.-J. Cho, B.-G. Park, H. Park, I. Martin, and C. D. Batista, Experimental confirmation of the magnetic ordering transition induced by an electronic structure change in the metallic triangular antiferromagnet Co1/3TaS2, Phys. Rev. B **113**, 125126 (2026).

[28] S. Mangelsen, P. Zimmer, C. Näther, S. Mankovsky, S. Polesya, H. Ebert, and W. Bensch, Interplay of sample composition and anomalous Hall effect in CoxNbS2, Phys. Rev. B **103**, 184408 (2021).

[29] H. Tanaka et al., Large anomalous Hall effect induced by weak ferromagnetism in the noncentrosymmetric antiferromagnet CoNb3S6, Phys. Rev. B **105**, L121102 (2022).

[30] J. M. Ziman, *Electrons and Phonons: The Theory of Transport Phenomena in Solids* (Oxford Univ. Press, 2001).

[31] P. Popčević et al., Electronic transport and magnetism in the alternating stack of metallic and highly frustrated magnetic layers in Co1/3Nb S2, Phys. Rev. B **107**, 1 (2023).

[32] J. Mi, T. Wang, W. Tang, X. Fan, J. Zhou, L. Zhang, L. Li, and Z. Xu, Chiral phonons and giant magneto-optical Raman effect in chiral antiferromagnetic CoNb3S6, Appl. Phys. Lett. **127**, (2025).

[33] H.-L. Luo et al., Discovery of Van Hove Singularities: Electronic Fingerprints of 3Q Magnetic Order in a van der Waals Quantum Magnet, ArXiv arXiv:2601.11796 (2026).

[34] J. Oh, M. D. Le, J. Jeong, J. Lee, H. Woo, W.-Y. Song, T. G. Perring, W. J. L. Buyers, S.-W. Cheong, and J.-G. Park, Magnon Breakdown in a Two Dimensional Triangular Lattice Heisenberg LuMnO3, Phys. Rev. Lett. **111**, 257202 (2013).

[35] J. Oh et al., Spontaneous decays of magneto-elastic excitations in non-collinear antiferromagnet (Y,Lu)MnO3, Nat. Commun. **7**, 13146 (2016).

[36] S. Hayami, R. Ozawa, and Y. Motome, Effective bilinear-biquadratic model for noncoplanar ordering in itinerant magnets, Phys. Rev. B **95**, 224424 (2017).

[37] S. Hayami and Y. Motome, Topological spin crystals by itinerant frustration, J. Phys. Condens. Matter **33**, 443001 (2021).

[38] Y. Akagi, M. Udagawa, and Y. Motome, Hidden Multiple-Spin Interactions as an Origin of Spin Scalar Chiral Order in Frustrated Kondo Lattice Models, Phys. Rev. Lett. **108**, 096401 (2012).

[39] H.-L. Kim, T. Saito, H. Yang, H. Ishizuka, M. J. Coak, J. H. Lee, H. Sim, Y. S. Oh, N. Nagaosa, and J.-G. Park, Thermal Hall effects due to topological spin fluctuations in YMnO3, Nat. Commun. **15**, 243 (2024).

[40] D. H. Lyons and T. A. Kaplan, Method for Determining Ground-State Spin Configurations, Phys. Rev. **120**, 1580 (1960).

[41] P. Park, W. Cho, C. Kim, Y. An, K. Iida, R. Kajimoto, S. Matin, S. S. Zhang, C. D. Batista, and J. G. Park, Spin Dynamics of Triple-Q Magnetic Orderings in a Triangular Lattice: Implications for Multi-Q Orderings in General Two-Dimensional Lattices, Phys. Rev. X **15**, 31032 (2025).

[42] D. Dahlbom et al., Sunny.jl: A Julia Package for Spin Dynamics, J. Open Source Softw. **10**, 8138 (2025).

[43] J. Bezanson, A. Edelman, S. Karpinski, and V. B. Shah, Julia: A Fresh Approach to Numerical Computing, SIAM Rev. **59**, 65 (2017).

[44] T. Hatanaka, J. Bouaziz, T. Nomoto, and R. Arita, Calculation of the Biquadratic Spin Interactions Based on the Spin Cluster Expansion for Ab initio Tight-binding Models, J. Phys. Soc. Japan **94**, 1 (2025).

[45] C. L. Henley, Ordering due to disorder in a frustrated vector antiferromagnet, Phys. Rev. Lett. **62**, 2056 (1989).

[46] F. Nickel, A. Kubetzka, S. Haldar, R. Wiesendanger, S. Heinze, and K. Von Bergmann, Coupling of the triple- q state to the atomic lattice by anisotropic symmetric exchange, Phys. Rev. B **108**, 1 (2023).

[47] T. Lu, Z. Feng, M. Ye, T. Kurumaji, and L. Ye, Elastoresistivity Signatures of Nematic Fluctuations in Layered Antiferromagnet CoTa3S6, 1 (2026): arXiv:2603.12704.

[48] U. Fano, Effects of configuration interaction on intensities and phase shifts, Phys. Rev. **124**, 1866 (1961).

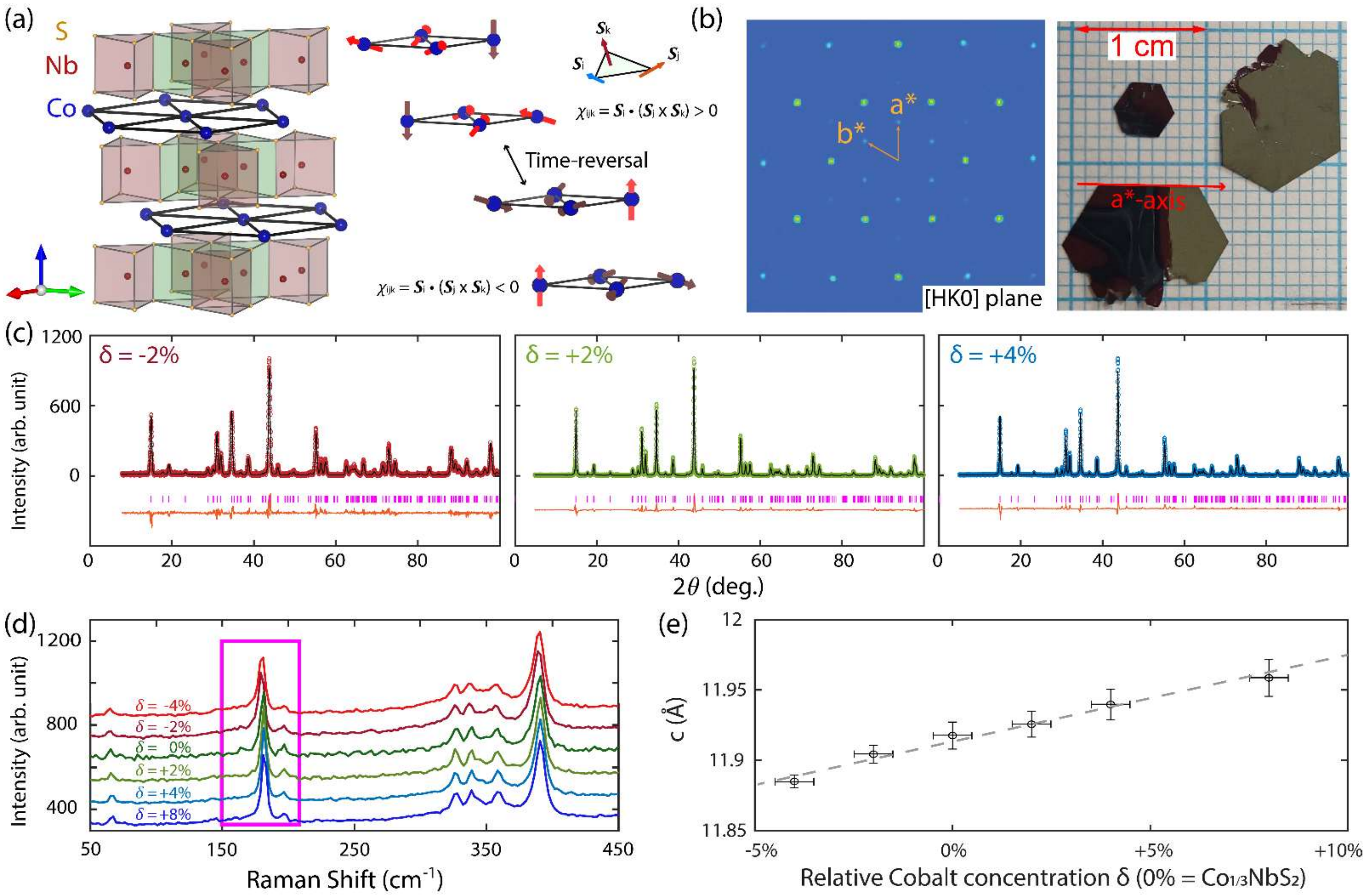

**Figure 1** (a) (Left) Crystal structure of $Co_{1/3}NbS_2$, where cobalt ions form a AB stacked triangular lattice inside the van der Waals gap of $NbS_2$ and (Right) Triple-**Q** magnetic structure which is responsible for spontaneous THE in this system (b) (Left) Precession image of [H K 0] plane obtained from single crystal X-ray diffraction, where cobalt superlattice peak observed, which is 30° rotated from the peak from $NbS_2$ structure and (Right) typical grown single crystal of clear hexagonal edges, which corresponds to $\boldsymbol{a}^*$-axis of $Co_{1/3}NbS_2$ (c) Powder X-ray diffraction and Rietveld refinement result with $Co_{1/3}NbS_2$ structure (d) Raman scattering result, where cobalt superlattice peak 150-200 cm$^{-1}$ is marked as pink rectangular box (e) Lattice spacing along $\boldsymbol{c}$-axis respect to cobalt concentration δ of $Co_{1/3(1\pm\delta)}NbS_2$, which linearly scales to the cobalt concentration variation.

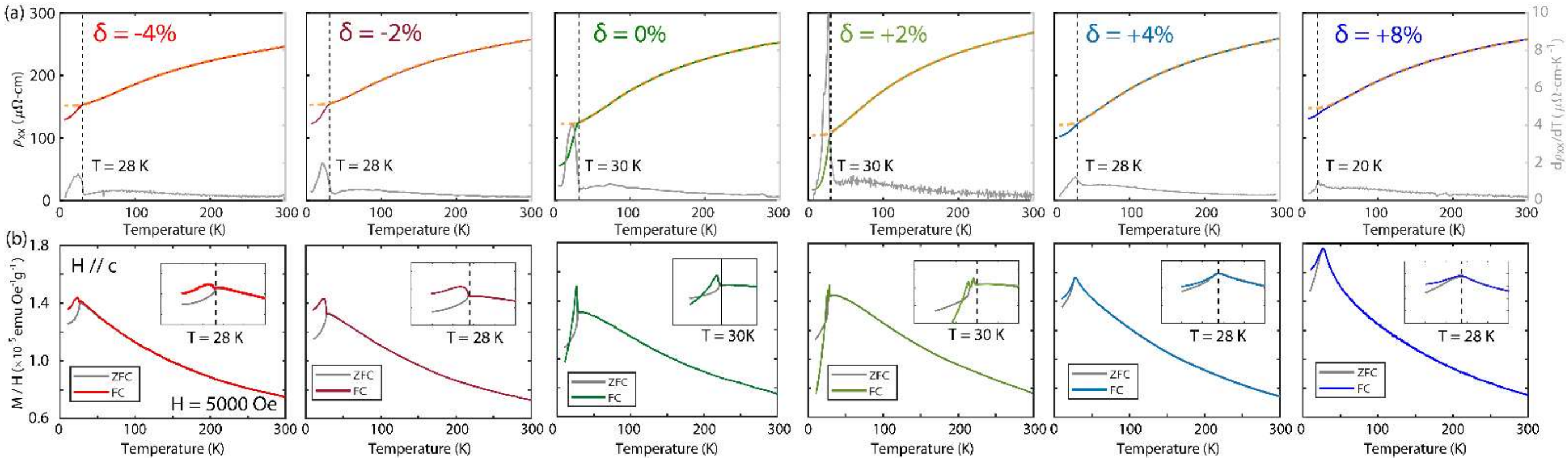

**Figure 2** (a) Temperature-dependent resistivity ($\boldsymbol{\rho_{xx}}$) along with the temperature derivative of resistivity (grey line) with respect to the variation of cobalt concentration. The orange dashed lines represent the fitting curves obtained from the phenomenological model discussed in the main text. (b) Magnetic susceptibility ($\boldsymbol{M/H}$) measured at 5000 Oe, zero-field-cooling (ZFC) and field-cooling (FC) measurements are overlaid to emphasize the distinct behavior observed for the composition which exhibits THE (from δ = -4% to +2% Co sample), in contrast to the similar behavior for ZFC and FC measurements in the Hall-inactive samples (δ = +4% and +8%).

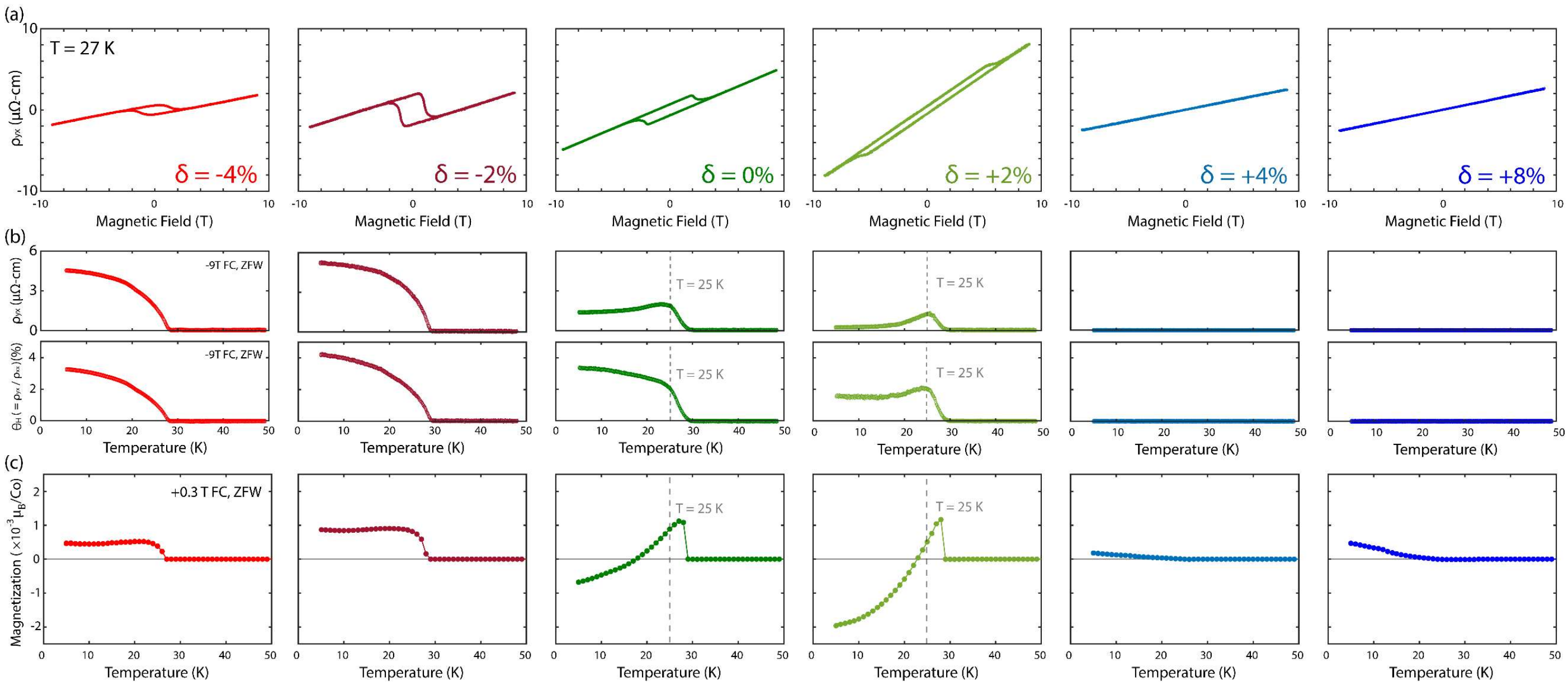

**Figure 3** (a) Field-dependent Hall resistivity $\rho_{yx}$ at T = 27 K for various cobalt compositions, highlighting the suppression of the topological Hall signal with increasing cobalt content. (b) Temperature dependence of the spontaneous Hall response plotted as (Top) Hall resistivity ($\rho_{yx}$) and (Bottom) Hall angle ($\theta_H = \rho_{yx}/\rho_{xx}$). The distinct anomaly observed near 25 K for δ = 0% and 2% samples is indicated as a grey dashed line. (c) Temperature dependence of the spontaneous magnetic moment with respect to cobalt composition.

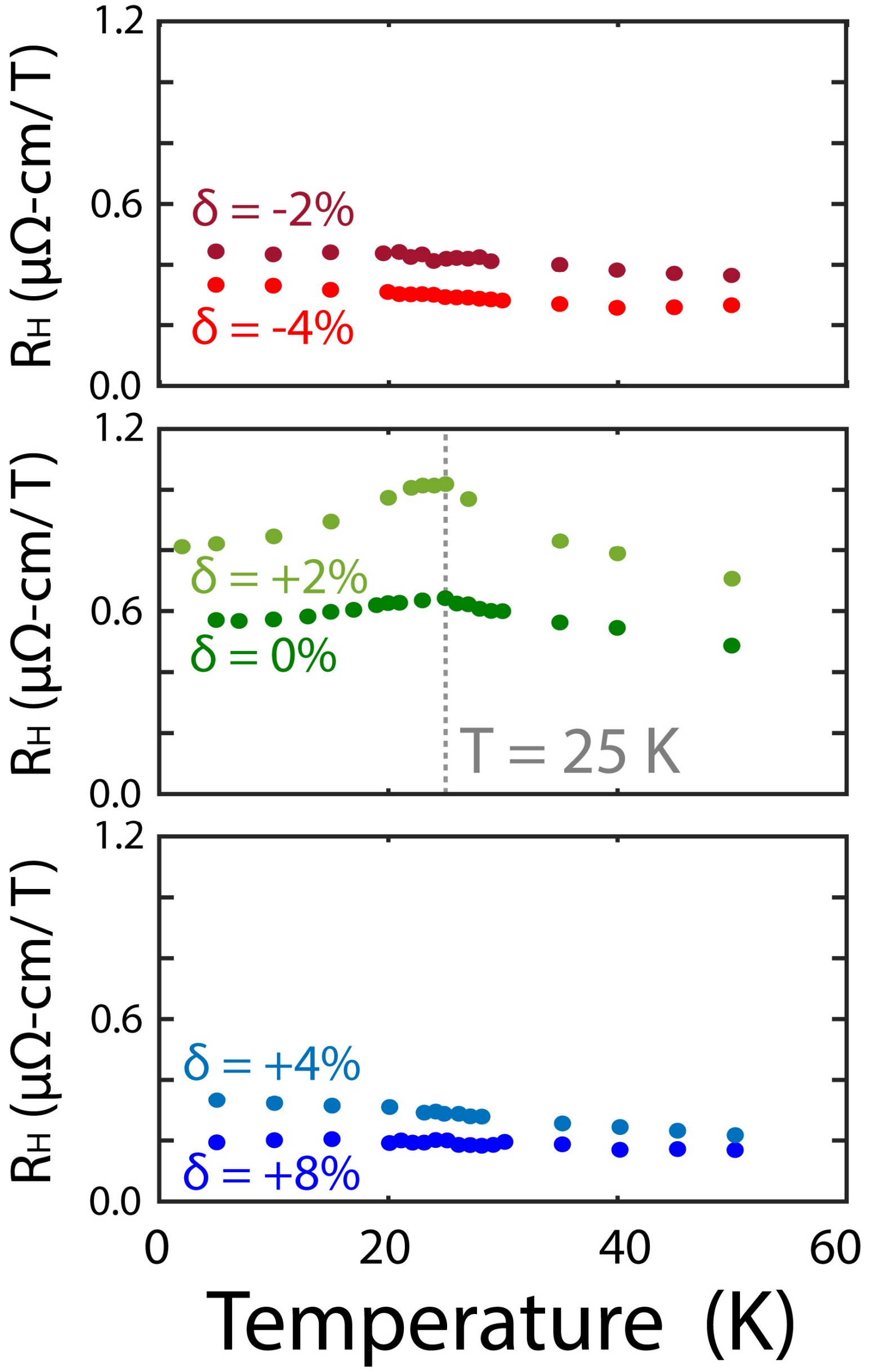

**Figure 4** Temperature dependence of ordinary Hall coefficient ($\boldsymbol{R_H}$) for each composition, (Top) δ = -4% and -2% sample (Middle) δ = 0% and 2% sample, (Bottom) δ = +4% and +8% sample. The grey dashed line indicates the peak behavior in $\boldsymbol{R_H}$ for δ = 0% and +2% sample, which is consistent with the peak position observed in the THE.

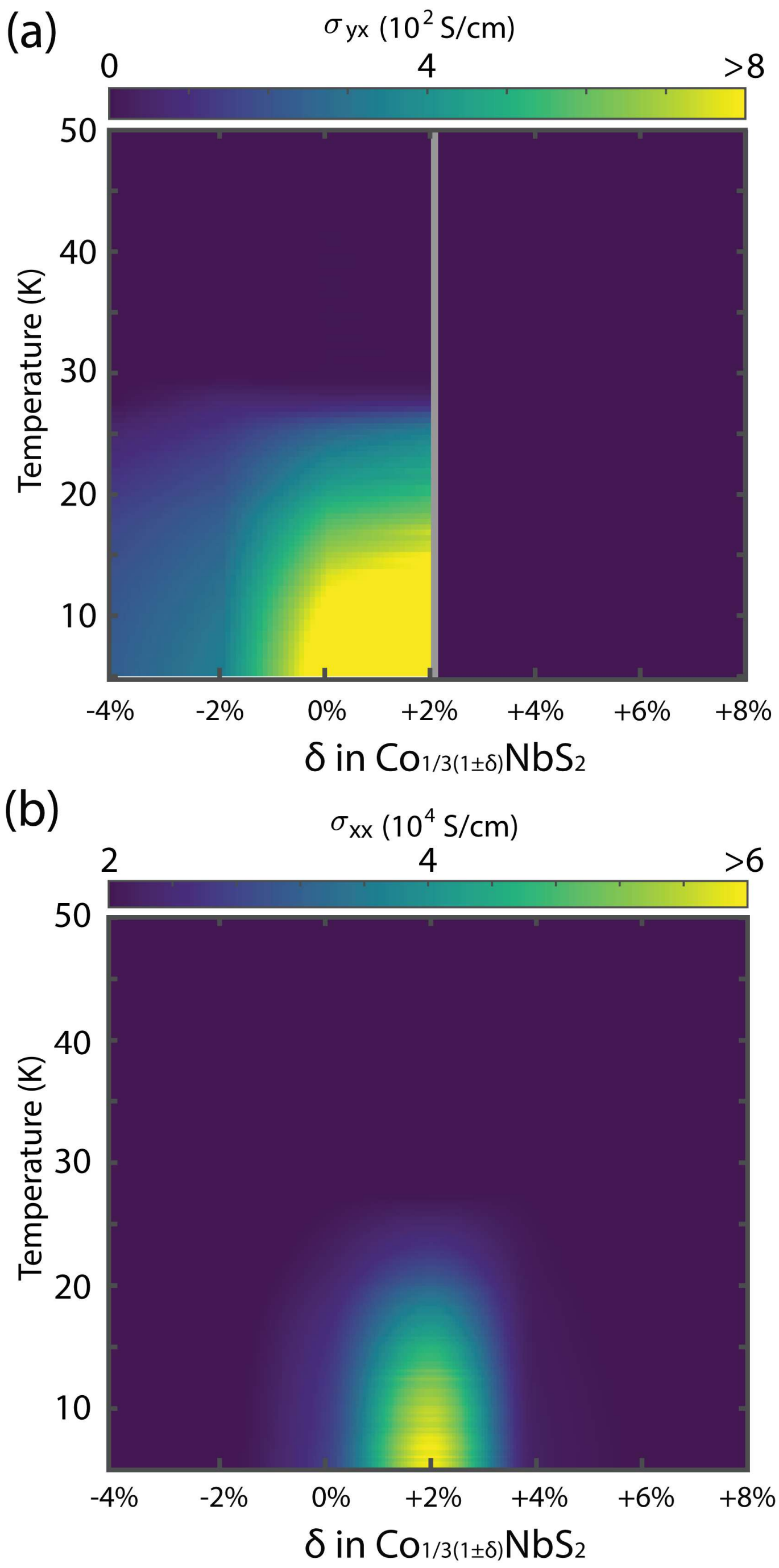

**Figure 5** (a) Transverse conductivity $\boldsymbol{\sigma_{yx}}$ of $Co_{1/3(1\pm\delta)}NbS_2$ at zero field as a function of Co composition and temperature. The THE is abruptly suppressed as Co content exceeds a critical threshold. (b) Longitudinal conductivity $\boldsymbol{\sigma_{xx}}$ of $Co_{1/3(1\pm\delta)}NbS_2$, which exhibits a distinct enhancement around δ = +2% composition.

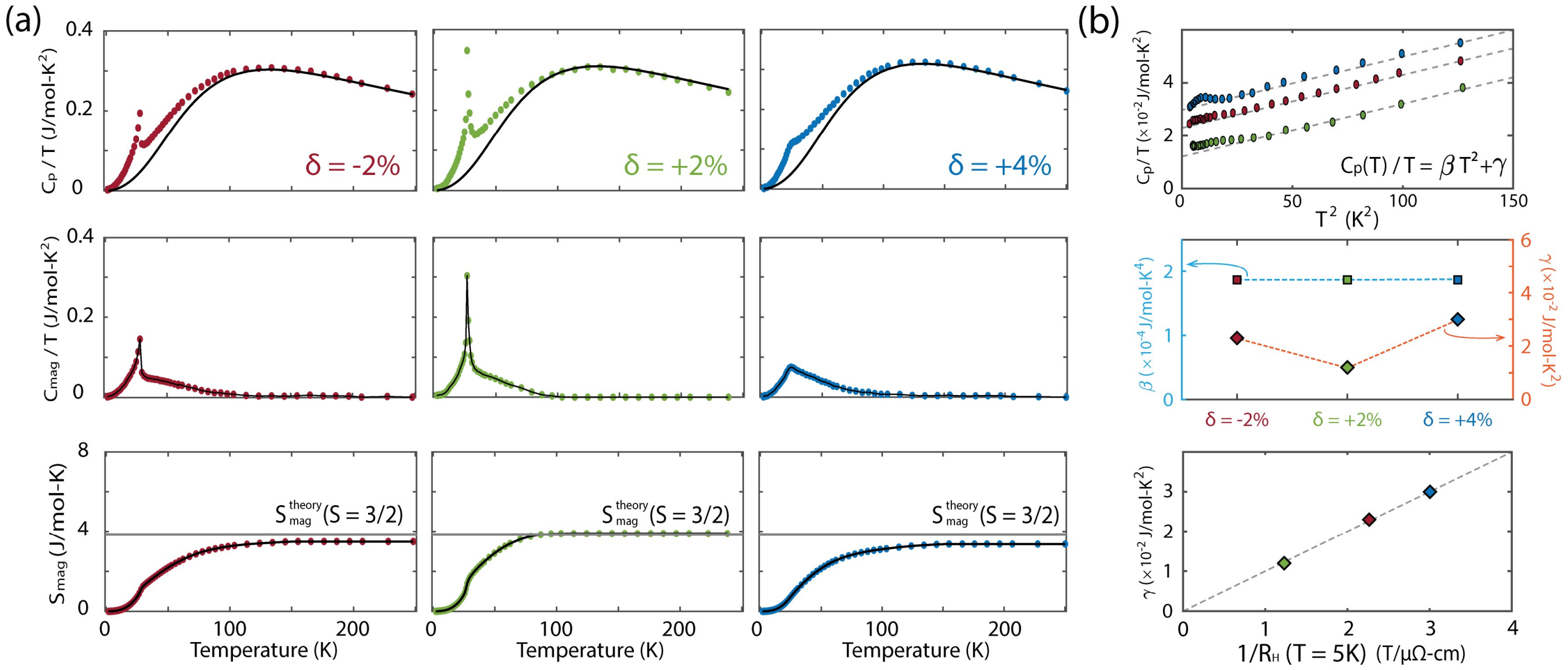

**Figure 6** (a) Temperature-dependent heat capacity divided by temperature ($C_p/T$) for δ = -2% (Left), +2% (Middle) and +4% (Right) samples. The magnetic contribution ($C_{\mathbf{mag}}/T$) and entropy ($S_{\mathbf{mag}}$) are isolated by subtracting the non-magnetic phonon and electron contributions. (b) (Top) Low-temperature heat capacity plotted as $C_p/T$ versus $T^2$, with linear fits used to extract the lattice ($\beta$) and electronic ($\gamma$) contributions. (Middle) Lattice contribution remains almost the same across the composition; electronic contribution shows a large variation across the composition. (Bottom) Sommerfeld coefficient scales linearly with the inverse of the ordinary Hall coefficient ($R_H^{-1}$) measured at 5 K.

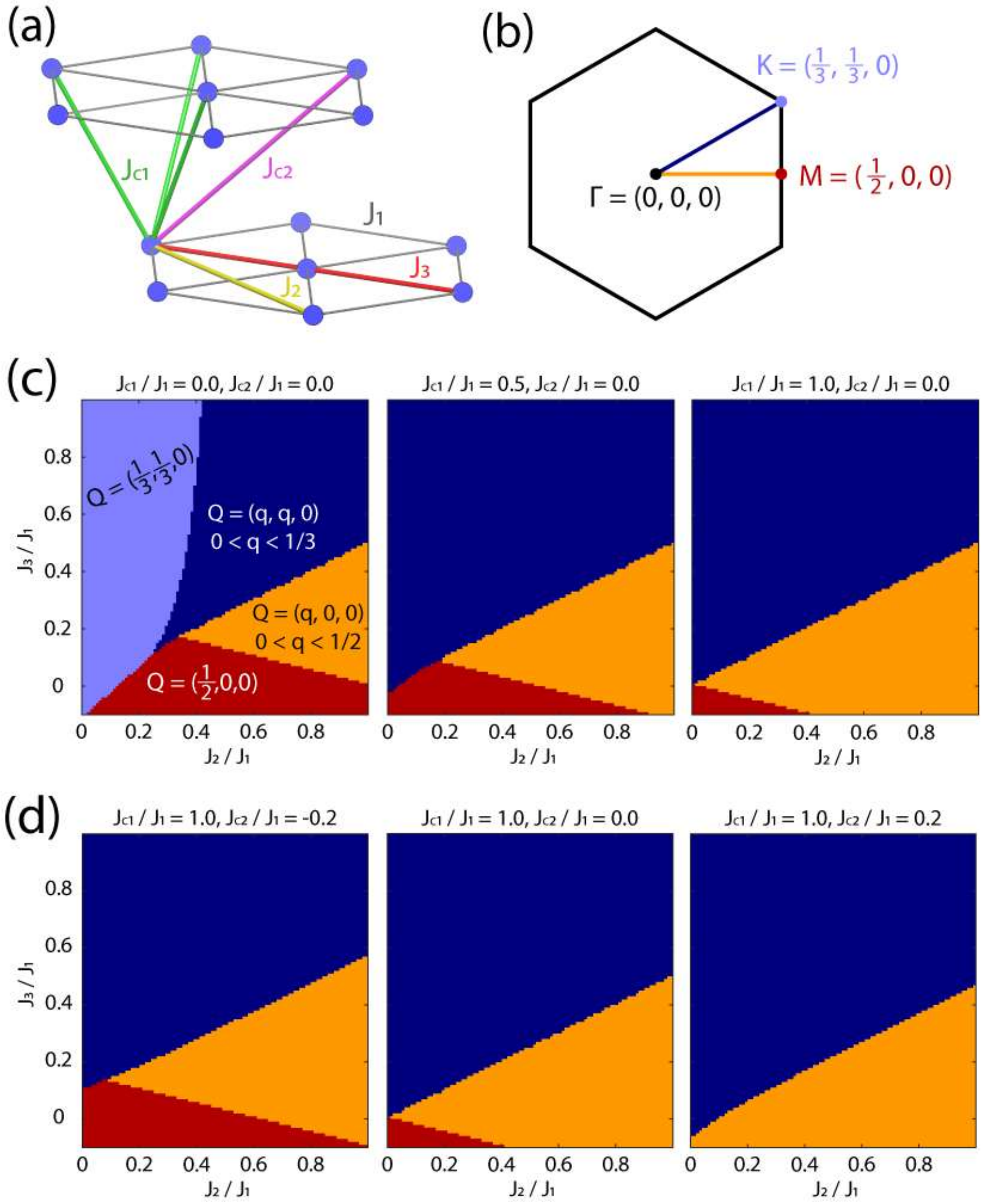

**Figure 7** (a) Schematic illustration of the exchange interaction paths ($\boldsymbol{J_1}, \boldsymbol{J_2}, \boldsymbol{J_3}, \boldsymbol{J_{c1}}, \boldsymbol{J_{c2}}$) for the Co-sublattice of $Co_{1/3}NbS_2$ structure. (b) The first Brillouin zone showing the high-symmetry points $\boldsymbol{\Gamma(0,0,0)}$, $\boldsymbol{K(1/3,1/3,0)}$ and $\boldsymbol{M(1/2,0,0)}$ corresponding to the modulation vectors analysed in (c). (c, d) Stable modulation vector for the spin Hamiltonian for $\boldsymbol{J_1 - J_2 - J_3}$ model ($\boldsymbol{J_1 > 0}$) for finite interlayer exchange interaction ($\boldsymbol{J_{c1}}, \boldsymbol{J_{c2}}$).

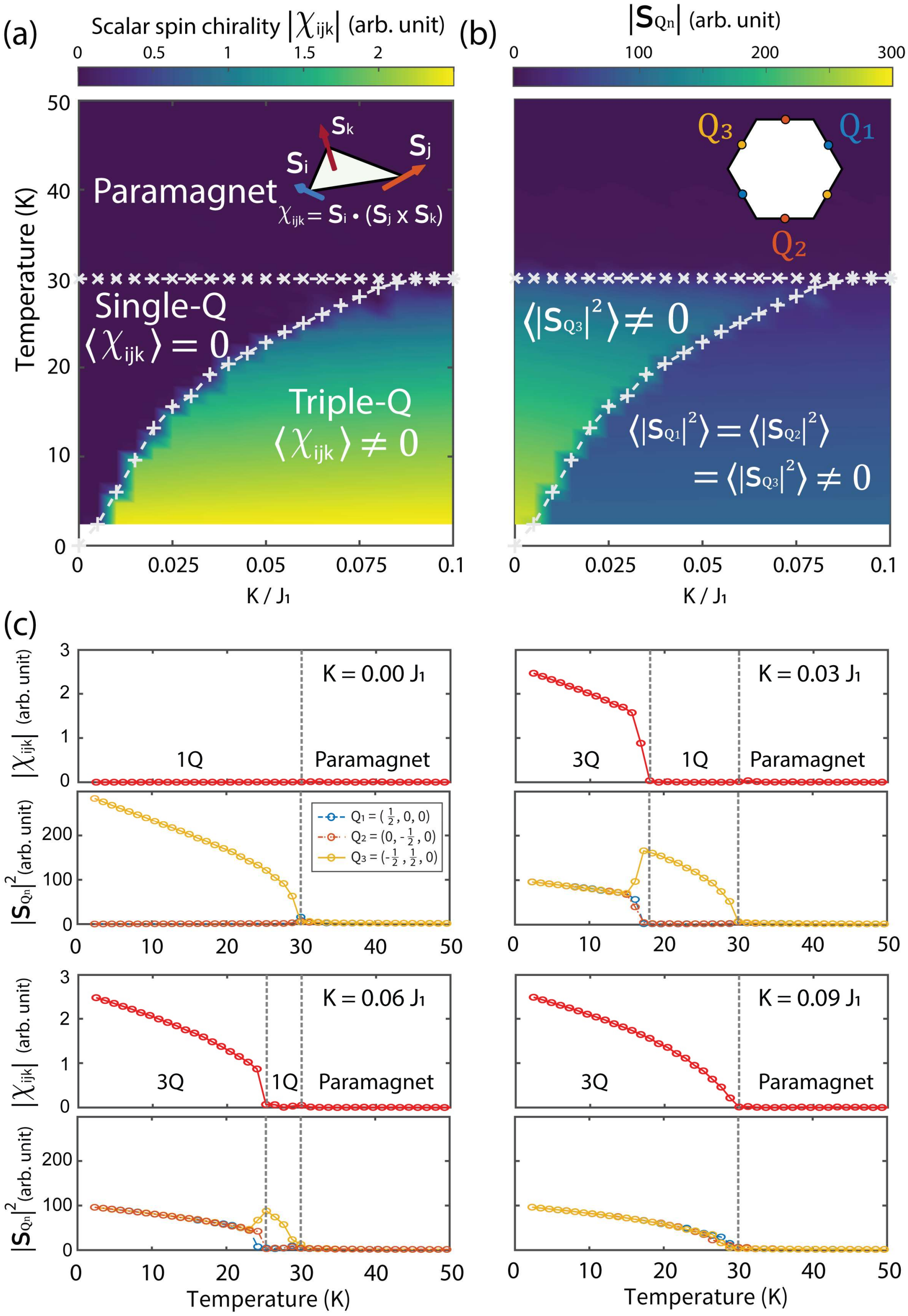

**Figure 8** Simulated finite temperature phase diagram for (a) spin scalar chirality and (b) square of Fourier component, which demonstrates the effect of biquadratic interaction in stabilizing triple-**Q** order as ground state. In terms of intensity of biquadratic interaction ($\boldsymbol{K}$), the phase boundary between paramagnet – single-**Q** – triple-**Q** changes and ultimately leads to paramagnet – triple-**Q** transition. (c) Temperature evolution of spin scalar chirality $\chi_{ijk} = \langle \boldsymbol{S_i} \cdot (\boldsymbol{S_j} \times \boldsymbol{S_k}) \rangle$, with the square of the Fourier component for the modulation vector ($\boldsymbol{Q_1}$ = (1/2, 0, 0), $\boldsymbol{Q_2}$ = (0, -1/2, 0) and $\boldsymbol{Q_3}$ = (-1/2, 1/2, 0)) for given biquadratic interaction intensity $\boldsymbol{K}$. Details are discussed in the main text.

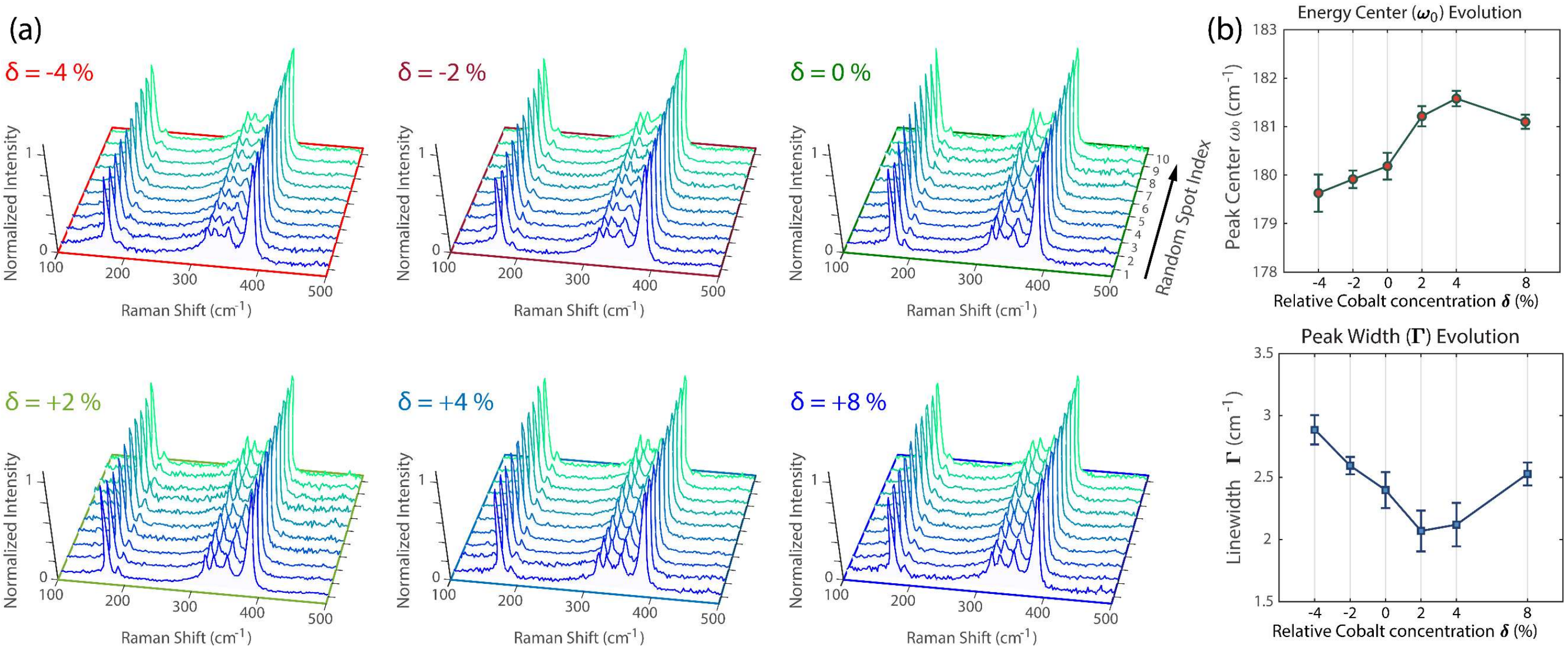

**Figure A1** (a) Raman spectra collected from ten randomly selected spatial locations on a representative single crystal. A linear background was subtracted from each spectrum, and the intensities were normalized to facilitate a direct spot-to-spot comparison of the line shapes, demonstrating macroscopic spatial homogeneity. (b) Compositional evolution of the phonon peak center ($\boldsymbol{\omega_0}$, the upper panel) and linewidth ($\boldsymbol{\Gamma}$, the lower panel) for the intercalation-induced mode emerging around 150 - 200 cm$^{-1}$ as a function of nominal Co concentration (δ). The data points and error bars denote the mean and standard deviation, respectively, of the parameters extracted via Fano line-shape fitting [48]

**Table 1.** Structural parameters obtained from Rietveld refinements of powder XRD of $Co_{1/3(1\pm\delta)}NbS_2$ at room temperature. Note that all refinements were performed using the $P6_322$ space group (No. 182).

| $\delta$ = -2%<br>$T$ = 300 K | Space group: P $6_3$ 2 2 (No. 182)<br>Cell dimension: $a = b$ = 5.7678(1) Å, $c$ = 11.8990(3) Å, $\alpha = \beta$ = 90°, $\gamma$ = 120° | | | | | |
|---|---|---|---|---|---|---|
| Atom | Wyckoff Position | x/$a$ | y/$b$ | z/$c$ | $B_{iso}$ (Å$^2$) | Occupancy |
| Nb | 2a | 0 | 0 | 0 | 1.54(1) | 0.16667 |
| Nb | 4f | 0.3333 | 0.6667 | 0.9983(2) | 1.55(6) | 0.33333 |
| Co | 2c | 0.3333 | 0.6667 | 0.25 | 1.10(1) | 0.16667 |
| S | 12i | 0.6696(2) | 0.0040(5) | 0.1338(2) | 1.37(5) | 1.00000 |
| | Agreement factors: $R_p$ (%) = 18.6, $R_{wp}$ (%) = 23.7, $R_{exp}$ (%) = 21.8, $\chi^2$ = 1.17 | | | | | |

| $\delta$ = +2%<br>$T$ = 300 K | Space group: P $6_3$ 2 2 (No. 182)<br>Cell dimension: $a = b$ = 5.7667(1) Å, $c$ = 11.9158(3) Å, $\alpha = \beta$ = 90°, $\gamma$ = 120° | | | | | |
|---|---|---|---|---|---|---|
| Atom | Wyckoff Position | x/$a$ | y/$b$ | z/$c$ | $B_{iso}$ (Å$^2$) | Occupancy |
| Nb | 2a | 0 | 0 | 0 | 0.98(6) | 0.16667 |
| Nb | 4f | 0.3333 | 0.6667 | 0.9985(2) | 0.28(4) | 0.33333 |
| Co | 2c | 0.3333 | 0.6667 | 0.25 | 0.18(9) | 0.16667 |
| S | 12i | 0.6682(9) | 0.0057(7) | 0.1331(2) | 0.51(4) | 1.00000 |
| | Agreement factors: $R_p$ (%) = 14.9, $R_{wp}$ (%) = 20.4, $R_{exp}$ (%) = 17.9, $\chi^2$ = 1.31 | | | | | |

| $\delta$ = +4%<br>$T$ = 300 K | Space group: P $6_3$ 2 2 (No. 182)<br>Cell dimension: $a = b$ = 5.7641(1) Å, $c$ = 11.9261(3) Å, $\alpha = \beta$ = 90°, $\gamma$ = 120° | | | | | |
|---|---|---|---|---|---|---|
| Atom | Wyckoff Position | x/$a$ | y/$b$ | z/$c$ | $B_{iso}$ (Å$^2$) | Occupancy |
| Nb | 2a | 0 | 0 | 0 | 0.67(6) | 0.16667 |
| Nb | 4f | 0.3333 | 0.6667 | 0.9986(1) | 0.39(4) | 0.33333 |
| Co | 2c | 0.3333 | 0.6667 | 0.25 | 0.52(8) | 0.16667 |
| S | 12i | 0.6707(8) | 0.0060(6) | 0.1329(1) | 0.43(4) | 1.00000 |
| | Agreement factors: $R_p$ (%) = 15.1, $R_{wp}$ (%) = 18.7, $R_{exp}$ (%) = 18.0, $\chi^2$ = 1.08 | | | | | |

**Table2**. Comparison between the nominal and experimentally measured Co/Nb atomic ratios for $Co_{1/3(1+\delta)}NbS_2$ single crystals. The relative composition δ, the corresponding scaling factors of the ideal stoichiometry (1+δ), and the nominal Co concentration x are listed alongside the average values determined by energy-dispersive X-ray (EDX) spectroscopy.

| Relative Composition δ | -4 % | -2 % | 0 % | +2 % | +4 % | +8 % |
|---|---|---|---|---|---|---|
| Stoichiometry Scale (1+ δ) | 96 % | 98 % | 100 % | 102 % | 104 % | 108 % |
| Nominal Co/Nb atomic ratio | 0.320 | 0.327 | 0.333 | 0.340 | 0.347 | 0.360 |
| Measured Co/Nb atomic ratio | 0.317 | 0.323 | 0.332 | 0.335 | 0.344 | 0.355 |

**Table 3.** Fitting parameters obtained from the extended Bloch-Grüneisen formula [see Eq. (2)] for each Co composition. To prevent overfitting and minimize parameter arbitrariness, the Debye temperature ($\Theta_D$ = 400 K) and the exponent ($n$ = 3) were fixed based on literature values [31]. The numerical regression was systematically performed within the high-temperature paramagnetic regime (50–300 K).

| Relative Composition δ | -4% | -2% | 0% | +2% | +4% | +8% |
|---|---|---|---|---|---|---|
| $\rho_0(\mu\Omega \cdot cm)$ | 159.73 | 158.31 | 125.24 | 104.21 | 123.38 | 147.45 |
| $\rho_1(\mu\Omega \cdot cm)$ | 53.35 | 49.99 | 20.35 | 49.38 | 55.17 | 64.36 |
| $\rho_2(\mu\Omega \cdot cm)$ | 125.55 | 140.00 | 193.17 | 215.89 | 177.81 | 133.20 |
| $\Theta_D(K)$ | 400 | 400 | 400 | 400 | 400 | 400 |
| $T_0(K)$ | 137.48 | 133.74 | 130.09 | 116.27 | 122.03 | 122.68 |